\documentclass[12pt]{article}
\pdfoutput=1
\usepackage{amssymb,amsmath,latexsym,times, mathtools, braket, physics, flexisym, setspace, amsmath, listings, graphicx, indentfirst, amsthm, braket}
\usepackage[margin=1in, footskip=0.25in]{geometry}
\newtheorem{theorem}{Theorem}[section]
\newtheorem{lemma}[theorem]{Lemma}

\begin{document}
\date{\today}
\author{Jupinder Parmar, Saarim Rahman, Jaskaran Thiara}
\setlength{\abovedisplayskip}{3pt}
\setlength{\belowdisplayskip}{3pt}

\title{A Formulation of a Matrix Sparsity Approach for the Quantum Ordered Search Algorithm}
\maketitle
\abstract{One specific subset of quantum algorithms is Grover’s Ordered Search Problem (OSP), the quantum counterpart of the classical binary search algorithm, which utilizes oracle functions to produce a specified value within an ordered database. Classically, the optimal algorithm is known to have a $\log_2 N$ complexity; however, Grover’s algorithm has been found to have an optimal complexity between the lower bound of $((\ln N-1)/\pi \approx 0.221\log_2 N)$ and the upper bound of $0 .433\log_2 N$. We sought to lower the known upper bound of the OSP. With [E. Farhi {\textit{et al}}, arXiv:quant-ph/9901059], we see that the OSP can be resolved into a translational invariant algorithm to create quantum query algorithm restraints. With these restraints, one can find Laurent polynomials for various $k$ -- queries -- and $N$ -- database sizes -- thus finding larger recursive sets to solve the OSP and effectively reducing the upper bound. These polynomials are found to be convex functions, allowing one to make use of convex optimization to find an improvement on the known bounds.  According to [A. Childs {\textit{et al}}, arXiv:quant-ph/0608161v1], semidefinite programming, a subset of convex optimization, can solve the particular problem represented by the constraints . We were able to implement a program abiding to their formulation of a semidefinite program (SDP), leading us to find that it takes an immense amount of storage and time to compute. To combat this setback, we then formulated an approach to improve results of the SDP using matrix sparsity. Through the development of this approach, along with an implementation of a rudimentary solver, we demonstrate how matrix sparsity reduces the amount of time and storage required to compute the SDP --- overall ensuring further improvements will likely be made to reach the theorized lower bound.

\section{Background}
The implementation of quantum computing, which would increase computational speed, has long been sought after as a replacement to the classical computer. This proposed advantage arises through the ability of quantum computers to utilize quantum mechanical principles of superposition and entanglement. Both of these properties allow quantum bits to store information that cannot be stored in the classical bit, leading to the ability of quantum computers to calculate multiple processes simultaneously. Given this capability, quantum computers can solve algorithms, such as Grover's Ordered Search Problem (OSP), with fewer queries than their classical analog [10]. The OSP's purpose is to essentially search through an ordered list in order to find the find a specified item. \par

The quantum OSP has an algorithmic complexity improvement over its classical counterpart; however, as the final decreased complexity value has not been formally stated, there are accepted upper and lower bounds. Once these two bounds are proven to be the same value, then the final algorithmic improvement will be known, and the OSP will be at its optimal state. Because $((\ln N-1)/\pi \approx 0.221\log_2 N)$ is currently the lowest known algorithmic complexity of the OSP, the problem lies in lowering the upper bound so that the difference between the two bounds is minimized. In order to do so, we seek to construct a quantum algorithm that finds an item in an ordered list of size $N$, with at most $k$ queries, in such a way that $N$ is maximized and $k$ is minimized (we will continue the use of $N$ and $k$ in this sense throughout our paper). Quantum algorithms for larger values of $N$ can specifically be found by applying a smaller case of ($k$, $N$), which has been proven to exist, recursively.

However, the process of lowering the upper bound has stagnated in recent years, as the methods of optimization that have been proposed require extensive amounts of time and storage to solve. Thus, to combat this obstacle, one can effectively find a translationally invariant algorithm (TIA) which represents the OSP in classical methods \cite{1}. With the TIA a quantum query algorithm can be solved and characterized by polynomials. These polynomials follow a convex pattern which can be exploited through semidefinite programming, a convex optimization technique \cite{2}. As found by earlier papers, the lowest known upper bound when using this method has been $4\log_{605} N \approx 0.433\log_2 N$ \cite{2}. However, we noticed that for cases within the SDP where $N$ is in the hundreds, solving becomes problematic because the SDP consumes extensive amounts of memory, as it must process around $N^6$ amount of computations. To overcome this issue, we developed an approach for solving the SDP using matrix sparsity which will vastly decrease the amount of time and memory needed to compute a result. With this approach we developed a rudimentary sparse solver and proved that it takes less time than the previous, dense solver. This ensures that implementing a fully capable sparse solver will likely lead to further bound improvements. \par
The rest of our paper follows this general outline: creation of a translationally invariant algorithm (TIA), the formulation of a semidefinite program, and the development of a matrix sparsity approach within the semidefinite program.\par
	Throughout the paper we will use standard notation from quantum computing as stated within [8]. In particular if there is a $n$-dimensional Hilbert space, the calculational basis is given by $ \ket{i},  i=0, \ldots, n-1$. An introduction to Grover's algorithm can also be found in Section 1.7 of [8].

\section{Formulation of Translation Invariant Algorithm}
The following description details the formulation of a TIA quantum algorithm for the OSP as proposed by [1, 2]. As this method is instrumental to the rest of our paper, we will go into detailed explanation, while following [1, 2], to ensure sufficient understanding.\par
Within the standard query model of the OSP, a query to a specific index of the list outputs a result of whether the desired component is before or after the queried position. Formally this can be stated as: when the target item is at index $j \in \{0,1,\ldots, N-1\}$, its location can be defined in the function $f_j: \{0,1,\ldots, N-1\} \rightarrow \{\pm 1\}$ as

\begin{equation}
f_j (x) :=
\begin{cases}
-1 \quad x < j \\
+1 \quad x \geq j. \\
\end{cases}
\end{equation}

In order to reach the state of the OSP where it uses as few queries to $f_j$ as possible to produce our target item, symmetry within the OSP can be used as: for $j \in \{0,1,\ldots, N-2\}$, changing the target item from $j$ to $j + 1$, will produce:

\begin{equation}
f_{j + 1} (x) :=
\begin{cases}
-1 \quad x = 0 \\
f_j (x - 1)  \quad  1 \leq x < N, \\
\end{cases}
\end{equation}

As observed, the condition at position 0 must be treated differently from the rest of the list; however, one can overcome this difference by extending $f_j$ to the function $g_j:  \mathbb{Z}/2N \rightarrow \{\pm 1\}$ defined as

\begin{equation}
g_j (x) :=
\begin{cases}
f_j (x)  \quad 0\leq  x < N \\
-f_j(x - N)  \quad  N \leq x <  2N. \\
\end{cases}
\end{equation}

as $j \in \{0,1,\ldots, N-1\}$, and

\begin{equation}
g_j (x) := -g_{j-N} (x)
\end{equation}

for $j \in \{N,N + 1,\ldots, 2N-1\}$, where all computations are done in mod $2N$. By using this modified function the symmetry expressed in (2) now appears translationally equivariant in the group $\mathbb{Z}/2N$ without the boundary condition at 0, namely as

\begin{equation}
G_{j+\ell} (x) = g_j (x - \ell)  \quad \forall \ j, x, \ell \in \mathbb{Z}/2N
\end{equation}

Concisely, these formulations simply state that for finding a target value at an index in the list, a function $g_j$ can be deduced within the subset of all integers that will display translational equivariance. \par
For this problem, the functions $f_j$ and $g_j$ are equivalent because a single query to $f_j$ can be simulated by simply querying $g_j$ on the original value of $x \in \{0,1,\ldots, N-1\}$, and vice versa given by (3). Therefore, using the function $g_j$, which does not have any problematic boundaries, will ensure that the query complexity of the OSP remains the same.		
In the quantum mechanical version of the query model, access to the query function is provided by a unitary transformation of the phase oracle -- a device which acts as the black box function for the OSP -- for $g_j$, a linear operator $G_j$, defined on the computational basis states ${ \ket{x} :x \in \mathbb{Z}/2N}$ as:

\begin{equation}
G_j \ket{x} := g_j (x)\ket{x}
\end{equation}

A k-query quantum algorithm is specified by an initial quantum state $\ket{\psi_0}$ and a sequence of ($j$-independent) unitary operators $U_1, U_2,\ldots, U_k$ -- essentially a series of matrix transformations. The algorithm begins in the initial state and afterwards a series of  query transformations and unitary operators ($U_j$) are applied alternately, in order to give the final quantum state:

\begin{equation}
\ket{\phi_j} := U_kG_jU_{k-1}\ldots U_1G_j \ket{\psi_0}
\end{equation}

In common terms: as the beginning ket vector, unitary operators, and query transformation can be modeled by a matrix, the algorithm is essentially multiplying these various matrices together until a final one is obtained. The final matrix provides the end quantum state of the quantum bits, and thus provides our specified index as the result. \par
To ensure that the quantum algorithm is exact as given by $\braket{\phi_j}{\phi_{j\textprime}} = \delta_{j,j\textprime}$ for all $j, j\textprime \in \{0, 1, \ldots , N - 1\}$, our goal is for each value of $N$ to be $\ket{\phi_0}$ and $U_1, U_2, \ldots, U_k$ for $k$ as small as possible. By exploiting the translation equivariance (5) of the function $g_j$, a simple approach to finding a better quantum algorithm for the OSP can be found. Simply, the equivariance can be described as a symmetry of the query operators in terms of the translation operator $T$ defined by:

\begin{equation}
T\ket{x} := \ket{x \ + \ 1} \quad \forall j \in \mathbb{Z}/2N
\end{equation}

we have

\begin{equation}
TG_jT^{-1} =  G_{j+1} \quad \forall x \in \mathbb{Z}/2N
\end{equation}

Thus the initial state can be given by:

\begin{equation}
\ket{\psi_0} = \frac{1}{\sqrt {2N}} \sum_{x=0}^{2N-1} \ket{x}
\end{equation}

Which satisfies $T\ket{\psi_0} = \ket{\psi_0}$  and the unitary operators abiding to

\begin{equation}
TU_tT^{-1} =  U_t
\end{equation}

for $t  \in \{1, 2, \ldots ,k\}$. In addition (9) holds for all $j \in \mathbb{Z}/2N$, ensuring that $j \in \{0,1,\ldots,N - 1\}$.
Fittingly, the $N$ possible orthogonal final states can mark the index of the targeted item as follows:

\begin{equation}
\ket{\phi_j} :=
\begin{cases}
\frac{1}{\sqrt 2}(\ket{j} \ + \  \ket{j + N}) \quad k \  even \\
\frac{1}{\sqrt 2}(\ket{j} \  -  \  \ket{j + N}) \quad k \  odd\\
\end{cases}
\end{equation}

A TIA is formed when an algorithm abides to (10), (11) and (12) and for an ordered search this TIA can find our marked item if $j = 0$ then it will be found for all $j \in \mathbb{N}$. With (9), $T^{-j}G_jT^j = G_0$. Providing us with:

\begin{equation}
\ket{\phi_j} = (T^j U_k T^{-j}) G_j \ldots (T^jU_1T^{-j}) G_j (T^j \ket{\psi_0})
\end{equation}

\begin{equation}
= T^j U_k (T^{-j} G_j T^j) U_{k-1} \ldots U_1 ( T^{-j} G_j T^j) \ket{\psi_0}
\end{equation}

\begin{equation}
= T^j U_k G_0 U_{k-1} \ldots U_1 G_0 \ket{\psi _0}
\end{equation}

\begin{equation}
= T^j \ket{\phi_0}
\end{equation}

Due to the nature of the TIA, it can easily be formulated as Laurent polynomials which are functions $Q : \mathbb{C} \rightarrow \mathbb{C}$ that can be written as:

\begin{equation}
Q(z) = \sum_{i= -D}^{D} q_iz^i
\end{equation}

for some positive integer D, where D is the degree of $Q(z)$. $Q(z)$ is nonnegative and symmetric iff $\abs{z} = 1$, $Q(z) \geq 0$, and $Q(z) = Q(z^{-1})$ for all $z \in \mathbb{C}$. For example, if $q_i = q_{-i}$ for all $i \in \{1, 2, \ldots, D\}$, simply meaning that $Q(z)$ is nonnegative and symmetric iff $q_i = q_{-i}  \in  \mathbb{R}$ for all $z \in \{0, 1, \ldots, D\}$. A representation of one such symmetric Laurent polynomial for the OSP is a Hermite Kernel (a sequence of orthogonal polynomials) of degree $N - 1$. The symmetricity and nonnegativity of $Q(z)$ in needed to ensure that $Q(z)$ abides to the conditions of the TIA.

\begin{equation}
H_N(z) := \sum_{i= -(N-1)}^{N - 1} \left(1 - \frac{|i|}{N}\right)z^i
\end{equation}

\begin{equation}
= \frac{1}{N} \left(\frac{z^{-N} - 1}{z^{-1} -1}\right)\left(\frac{z^N - 1}{z - 1}\right).
\end{equation}

The following, given by [1], characterizes the TIA in the form of Laurent Polynomials.
\begin{theorem}[{[3]}]
There exists an exact, translation invariant, k-query quantum algorithm for the N-element OSP if and only if there exist nonnegative, symmetric Laurent polynomials $Q_0(z),...,Q_k(z)$ of degree $N - 1$ such that
\end{theorem}

\begin{equation}
Q_0(z) = H_N(z)
\end{equation}

\begin{equation}
Q_t(z) = Q_{t-1}(z) \quad at \quad z^N = (-1)^t \quad \forall  \ t \in \{1,2,\ldots,k\}
\end{equation}

\begin{equation}
Q_k (z) =1
\end{equation}

\begin{equation}
\frac{1}{2\pi} \int_0^{2\pi} Q_t (e^{iw}) \ dw =1 \quad \quad \forall t \in \{0, 1, \ldots, k\}.
\end{equation}

Each polynomial $Q_t(z)$ represents the quantum state of the OSP after $t$ queries. Given by

\begin{equation}
Q_t(z) = \sum_{i= -(N-1)}^{N - 1}q_i^{(t)} z^i
\end{equation}

Which follows:

\begin{equation}
q_i^{(t)} = 2\sum_{m=1}^{N - i}\braket{\psi_t}{N - m}\braket{N - m - i}{\psi_t}
\end{equation}

For:

\begin{equation}
\ket{\psi_t} := U_t G_0 U_{t-1} \ldots U_1 G_0 \ket{\psi_0}
\end{equation}

Using polynomials satisfying (20 - 23)  one can reconstruct all unitary operators for the OSP using (25), thus proving that set of $(k,N)$ is a solution to the algorithm.

\section{Optimization Techniques}
The given restraints (20 - 23) create various Laurent polynomials which can be seen to abide by the conditions of a convex function, essentially a continuous function whose second derivative is positive, giving: $f: X \rightarrow \mathbb{R}$, with $X$ being a space of variables. With these constraints one can use convex optimization to find TIA's for various cases until the highest value of $N$ for a given $k$ is discovered. When searching for a convex optimization technique, there are various drawbacks in the following methods: conjugate gradient descent, simplex method, and zero temperature annealing.  Conjugate gradient descent works to solve and optimize problems with a given cost function, while ours does not necessarily contain one, leading to solving inefficiencies. The simplex method was not satisfactory as, for larger problems, it produces a higher probability of error and time consumption due to a large amount of operations. In addition, both of the previous methods do not utilize knowledge about the  absence of local minima in their search. Zero-temperature annealing is more optimal; however, it performs poorly because the direction of movement is random [7]. Although none of the previous methods were a perfect fit for our project, semidefinite programming is able to solve both of our needed constraints, thus we proceeded with the method which is formulated in [2].

\section{SDP Formulation}
In order to apply semidefinite programming to the OSP, one would use the spectral factorization of nonnegative Laurent polynomials to create a set of conditions based on equations (20 - 23) as linear constraints. Effectively producing  semidefinite matrices which can be diagonalized. This method follows from the Fejer-Riesz theorem, which essentially states that for a function  $f:\mathbb{R} \rightarrow \mathbb{C} $, the Cesaro mean of the fourier series of $f$ converges uniformly to $f$ on $[-\pi,\pi]$: \par

\begin{theorem}[{[4,5]}]
Let $Q(z)$ be a Laurent polynomial of degree D. Then $Q(z)$ is nonnegative if and only if there exists a polynomial $P(z) = \sum_{i = 0}^{D} p_iz^i$ of degree D such that $Q(z) = P (z)P (1/z∗)∗$. Let $Tr_i$ denote the trace along the $i$th super-diagonal (or ($−i$)th sub-diagonal, for $i < 0$), i.e., for an $N \times N$ matrix $X$
\end{theorem}

\begin{equation}
Tr_iX =
\begin{cases}
\sum_{\ell = 1}^{N - i} X_{\ell,\ell + 1} \quad i \geq 0\\
\sum_{\ell = 1}^{N + i} X_{\ell - i, \ell} \quad i < 0\\
\end{cases}
\end{equation}

The following lemmas can be explicitly deduced from [2], we will simply outline each.
		 	 	 		
\begin{lemma}			
Let $Q(z) = \sum_{i = - (N - 1)}^{N - 1}q_iz^i$ be a Laurent polynomial of degree $N - 1$. Then $Q(z)$ is nonnegative if and only if there exists an $N \times N$ Hermitian, positive semidefinite matrix $Q$ such that $q_i = Tr_i Q$.
\end{lemma}

Lemma 4.2 is simply stating a way in which to prove that a given Laurent polynomial is nonnegative. Then with Lemma 4.3:

\begin{lemma}		 					
If $Q(z)$ is a nonnegative, symmetric Laurent polynomial, then the matrix $Q$ in Lemma 4.2 can be chosen to be real and symmetric without loss of generality.
\end{lemma}

One can show that these Laurent polynomials are also symmetric.

With these Lemmas, we can build upon conditions (20 - 23) to create the SDP [2]. The SDP: $(S(k,N)$ will find real symmetric positive semidefinite $N \times N$ matrices $Q_0, Q_1,\ldots, Q_k$ satisfying:

\begin{equation}
Q_0 = E/N
\end{equation}

\begin{equation}
\mathcal{T}_t Q_t = \mathcal{T}_tQ_{(t-1)} \quad \forall t \in \{1,2,\ldots,k\}
\end{equation}

\begin{equation}
Q_k = I/N
\end{equation}

\begin{equation}
TrQ_t = 1 \quad \forall t \in \{0,1,\ldots,k\}
\end{equation}

where $E$ is the $N \times N$ matrix in which every element is 1 and $\mathcal{T}_t : \mathcal{S}^N \rightarrow \mathbb{R}^{N - 1}$ is a linear operator  that computes signed traces along the off-diagonals, namely

\begin{equation}
(\mathcal{T}_tX)_i = Tr_iX + (-1)^tTr_{i - N}X
\end{equation}

Continuing our notion from before, Theorem 3 from [2] shows that if there is a solution to $(S(k,N))$ then there will be a given translationally invariant $k$-query quantum algorithm for the $N$-element OSP and vice versa.
The SDP will either guarantee a solution or find that none are feasible for the given proposal, $(k,N)$ using one of several available software packages [6].
By solving the SDP, a quantum query algorithm for a particular $N$ and $k$ can be found which lowers the upper bound for the quantum query complexity of the OSP.

\section{SDP Implementation}
Previously, we identified how a SDP can be applied to the OSP; however, it is not intuitive how this formulation can be implemented. Thus, in this section we outline how to do so through the solving of a small sample case where $k = 2$ and $N = 3$. After demonstrating our simple example of the SDP, we will illustrate how we programmed it to deal with larger cases.
In order to prove $(S(2,3))$ contains a valid solution, we will use the constraints (28 -  31) to find a semidefinite matrix $Q_1$. From (28), it is seen that $Q_0$ equals  $E/N$, thus:

\begin{equation}
\frac{1}{3}\
\begin{bmatrix}
1 & 1 & 1 \\
1 & 1 & 1 \\
1 & 1 & 1 \\
\end{bmatrix}
\end{equation}

From (30) we can deduce matrix $Q_2$ to be:

\begin{equation}
\frac{1}{3}
\begin{bmatrix}
1 & 0 & 0 \\
0 & 1 & 0 \\
0 & 0 & 1 \\
\end{bmatrix}
\end{equation}

To find our solution for matrix $Q_1$, we can take advantage of the proposed symmetry and populate it with the six variables: $q_{11}$, $q_{22}$, $q_{33}$, $q_{21}$, $q_{31}$, $q_{32}$ as shown below:

\begin{equation}
Q_1 =
\begin{bmatrix}
q_{11} & q_{21} & q_{13} \\
q_{21} & q_{22} & q_{32} \\
q_{31} & q_{32} & q_{33} \\
\end{bmatrix}
\end{equation}

However, due to (31), we can reduce the amount of variables we need by one as the main trace of $Q_1$ will equal one -- effectively eliminating the necessity for one main variable, say $q_{33}$.
Now we are left to find the values of: $q_{11}$, $q_{22}$, $q_{21}$, $q_{31}$, $q_{32}$ using (29) which utilizes the “signed” trace that [2] develops. Essentially, the signed trace will produce an equation in terms of these variables, and set them equal to the specified trace of either $Q_2$ or $Q_0$
As given by (29), we can use the two constraints:

\begin{equation}
\mathcal{T}_1Q_1 = \mathcal{T}_1Q_0 \quad \mathcal{T}_2Q_2 = \mathcal{T}_2Q_{1}
\end{equation}

To develop a set of equations with which to find the identities of our variables. One such example for our case would be:

\begin{equation}
\frac{1}{3} =  q_{21} + q_{32} - q_{31}.
\end{equation}

By assuming that one variable takes a proposed value, we can use the aforementioned equations to determine the rest of our variables. With this, our matrix $Q_1$ will be complete and as long as it is positive semidefinite, it will be proven that there exists an algorithm for the case. \par
When computing solutions for higher cases of the SDP, the implementation of a program is required as the amount of variables become extremely large. Thus, for the rest of this section, we outline how we used Python along with [6] to do so.
The SDP solver within [6], takes a set of parameters which consists of: a cost function, a pair of matrices which contribute to finding the variables of the needed matrices, and another two matrices which ensure the positive semidefiniteness of the proposed values in the matrices. Therefore, in our program we will need to set up these components for the solver to use.
First, we must determine the amount of variables that we need for the SDP.  As the matrices $Q_{1,2\ldots,N-1}$ are of size $N \times N$ and symmetric, we will only need the lower diagonal components, $N(N+1)/2$ entries. For the sake of simplicity we will present the number of variables as $\sigma$. Our program begins by displaying all these variables within a list, and due to its nature of converting between lists and matrices frequently, we defined a function to convert the position $(i,j)$ of a matrix into a list, using the following:

\lstset{language=Python}
\begin{lstlisting}
def list_position(i, j, k):
if (i < j):
  tmp = j
  j = i
  i = tmp
return N*(N+1)/2-(N-j)*(N-j+1)/2+i-j + (k-1)*N*(N+1)/2
\end{lstlisting}

Then, due to the lack of a cost function, we can simply define a zero vector with $\sigma$ components to ensure that the cost will always be zero.
\begin{lstlisting}
c = matrix(0.0, (num_variables(),1))
\end{lstlisting}
Note the function num_variables() returns the value of $\sigma$.

In order to create the linear constraint equations for the SDP we must use the signed trace formula of (32). We create it as follows:

\begin{lstlisting}
def signed_trace(A, t, k, i, row, fac=1.0):
for j in range (i, N):
   A[row, list_position(j,j-i,k)] = fac*1.0
   sign = fac*pow(-1.0,t)
   for j in range (N-i, N):
        A[row, list_position(j, j+i-N,k)] = sign
return
\end{lstlisting}

The constraint equations will then be supplied to a matrix, $A$ --- which will have $\sigma$ columns and $\rho$ rows, where $\rho$ represents the number of constraints -- where they will be econded into linear constraints. The solutions, or traces, for these equations will populate matrix $b$. These two matrices then allow us to formulate the equation: $A*x = b$, where $x$ is the column matrix that has the values for all the variables stored in it, allowing all our linear equations to be encoded. In our code, we explicitly define these two matrices as:

\begin{lstlisting}
A = spmatrix(0.0, [0], [0], ((k-1)+k*nrindep, num_variables()))
b = matrix(0.0, ((k-1)+k*nrindep,1))
\end{lstlisting}

Note “nrindep” equals the number of independent traces contained in the matrix: $N/2 -1$ if $N$ is even, and $(N-1)/2$ if $N$ is odd.
Finally, we must create the semidefinite condition within our SDP.  As given by the documentation from [6], the constraint can be stated as $Gs[k]*x + ss[k] = hs[k] \quad \forall k = 0, \ldots, N-1$. The solver will look for positive semidefinite matrices $Sk$ satisfying the above condition, so we must create our own matrices, $Gs_k$ and $Hs_k$. To ensure that our matrices $Q_{1,2\ldots,N-1}$ are semidefinite, it is enough to ensure that$ -Q_k + ss_k = 0$. We then must ensure that we choose a $Gs_k$ such that $G_{sk}$  gives matrix $Q_k$ where all the elements are written as a column vector in the right order. We do this by creating a semidefinite constraint function:

\begin{lstlisting}
def semidef_constraint(k):
G = spmatrix( 0.0, [0], [0], (N*N, num_variables()) )
for j in xrange (0,N):
  for i in xrange(j,N):
      G[i*N+j,list_position(i,j,k)] = -1.0
      G[j*N+i,list_position(i,j,k)] = -1.0
return G
\end{lstlisting}
We then use the constraint formula and this function in our creation of matrices $Hs$ and $Gs$.\par
Through the formulation of these objects, we will have satisfied all the necessary steps for the SDP solver. We output results as stated by:

\begin{lstlisting}
solution = solvers.sdp(c, A=A, b=b, Gs = Gs, hs = Hs)
\end{lstlisting}

\section{SDP Program Results}
Once we completed our program, we ran it for an extended amount of time on a 12 core desktop computer as well as an everyday Mac laptop. We found that it was infact, unfeasible to find bound improvements on the previously stated upper bound as, say for the case $k = 5$, $N = 5250$ we would need $N^6$ operations which is roughly $2.1 \times 10^{22}$. Due to each operation taking a clock cycle, it would take approximately $5250^6/32000000000 = 654,343,574,524$ seconds or more than 7 million days. Thus, one would need to create a more efficient method to compute results which utilizes less storage, computations, and time. We did however, verify smaller cases confirming the results of $(2, 6)$ and $(3, 56)$ being the largest cases for their respective amount of queries as found within [1,2].

\section{Matrix Sparsity}
As we showed earlier, finding an upper bound improvement for the OSP with currently available SDP solvers is impractical. This problem arises within [6] and other similar programs, as their KKT solver (the function which is essential to solving the SDP) converts sparse matrices into dense matrices. The drawback lies within the fact that sparse matrices store only the non-zero elements within its memory, while dense matrices store all elements, regardless of their value.  This function leads to an extensive amount of storage being taken up, which not only increases the time it takes to compute the program, but hinders it from computing solutions to higher values. Since our SDP uses matrices which are very sparse, sparse matrices would be a great tool. Thus, within this section we illustrate the superiority of a sparse solver for the SDP and our formulation of a rudimentary one, which currently works for small cases.\par

To quantify creating a sparse solver, we must demonstrate that the matrices we deal with in the constraint equations: $A*x = b$ and $Gs[k]*x + ss[k] = hs[k]$  are indeed sparse. We will begin our discussion by looking at matrices $A$ and $b$, then move onto matrix $Gs$.\par

Beginning with matrix $A$, we will first find the number of columns. It is seen that each $Q_i$ contains $\sum_{i=1}^{N}{i}$ variables, which is also equal to $N(N + 1)/2$. And, as $Q_0$ and $Q_k$ are known, there are  $k - 1$ matrices remaining, thus leading to the number of columns: $(k - 1) \times N(N + 1)/2$. Now we have to find the number of rows which each contain a single constraint, for which there are three cases. Case one occurs when $TrQ_i = 1$ and there are $(k -1)$ such constraints, hence there will be $(k - 1)$ rows. Since this case only uses the diagonal elements of $Q_i$, each row contains $N$ non-zero elements. Therefore, $\Omega$, which we use to symbolize the number of nonzero elements, for this case is equal to: $(k -1)(N)$. Case 2 consists of the signed traces at the boundary conditions: $\mathcal{T}_1Q_1 = \mathcal{T}_1Q_0$ and $ \mathcal{T}_kQ_k = \mathcal{T}_kQ_{k - 1}$. Due to knowing $Q_0$ and $Q_k$, we need only $N$ variables in each constraint, thus the question is how many constraints are contained within $\mathcal{T}_kQ_k = \mathcal{T}_kQ_{k - 1}$. Within this equation, we see that both sides are vectors of length $N - 1$; however, they are not all independent because of the symmetry of the $Q_k$’s. This leaves us with $N/2 - 1$ independent equations if $N$ is even, and $(N - 1)/2$ if $N$ is odd, we will refer to the number of independent equations from here on out as: $\Pi$. As there are two signed traces, the number of rows within this case is equal to $2*\Pi$. With each row needing $N$ variables, we see that $\Omega = N*2*(\Pi)$. Finally we look at  Case 3 which focuses on the the $(k -2)$ remaining intermediary signed traces. As stated before, the number of linear equation within each constraint is equal to $\Pi$, hence we need a total of $(k -2)*(\Pi)$ rows. Since now both the left boundary and right boundary of the signed trace equation are unknown, we will need $2N$ variables instead of $N$. Thus, for case 3, $\Omega = (k - 2)*(2N)*(\Pi)$.Therefore, we can conclude that $A$ has $(k-1) N(N +1)/2$ columns and $(k - 1) + 2(\Pi) + (k - 2)(\Pi)$ rows, meaning a total size of $(k -1)[(k -1) + k\Pi] * N(N + 1)/2$ for matrix $A$. Within $A$, we then see there to be a total of: $(k - 2)*(2N)*(\Pi) +  N*2*(\Pi) +  (k -1)(N)$ nonzero elements.

\begin{figure}
\includegraphics[width=\textwidth]{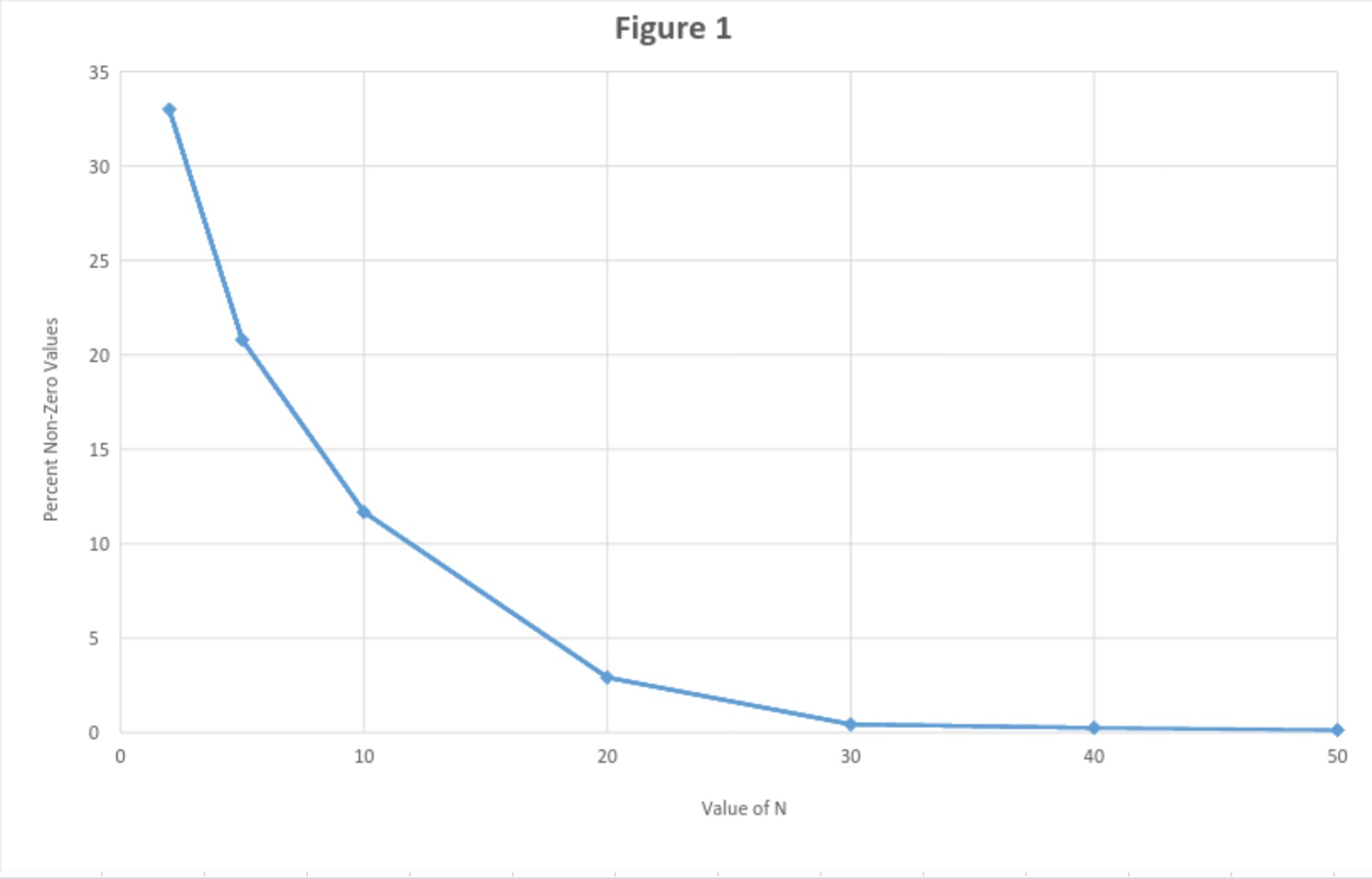}
\end{figure}

As Figure 1 shows, for the small case where $k = 3$, when the values of $N$ gets progressively larger, the percent of non zero elements immensely decreases, making it highly advantageous to use sparse matrices. For example,when  $k$ = 4 and $N$ = 250, sparse matrices would only store 186,750 entries in memory, while dense matrices will store 2,918,157,375 entries. Amazingly, the sparse matrix only needs stores $6.40 \times 10^{-5}$ percent of the values that the dense one does.\par
	For vector $b$ in the equation, we found that because it is storing the solutions to the equations in $A$, it will have the same number of rows, meaning: $R = N(N+1)/N$. Being a vector, $b$ will only have one column. The amount of $\Omega$ in $b$ can be found to equal: $[ \lceil N/2 \rceil + (k-2)]$, showing that the size of vector  $b$ will increase far more rapidly than the number of nonzero elements -- ensuring the necessity for sparsity.\par
	Next, our discussion will transition to the condition regarding semidefiniteness, as stated in: $Gs[k]*x + ss[k] = hs[k] \quad  \forall k = 0, \ldots, N-1$. As we did with the matrices $A$ and $b$, we will analyze the matrix $Gs$. Simply, the matrix $Gs$ will contain a set of matrices, $G$, within it. Each of these matrices will deal with the semidefinite constraints for their respective $Q_{1,\ldots,k-1}$ matrix, consequently ensuring that there will be $(k-1)$ $G$ matrices. Thus, due to $G$ populating $Gs$, one needs only to look at the sparsity of each $G$ to determine the sparsity of $Gs$ as a whole. With this in mind, we find that the number of rows within $G$, follows as: $R = N^2$; while the number of columns are defined by: $(k-1)\sum_{i = 1}^{i = N}i$ which is equal to $(k - 1) \times N(N +1)/2$. These formulations follow the existing logic found when describing $A$ and $b$; however, the difference between them lies within the fact that we are now dealing with semidefinite constraints. This difference results in matrices of $G$ generally having higher amount of $\Omega$, given by: $N^2$. We see that $\Omega$ follows this value as each $G$ has $N^2$ rows, and with each row encoding a constraint for its respective $Q_k$ matrix, it leads to there being $N^2$ non-zero elements.  Building upon our previous discussion of the case $(4,250)$, we see that there will be 187,5000 nonzero elements within $Gs$, while the combined $G$’s will have a total size of 5,882,812,500 elements, displaying the fact that nonzero elements will only take up  $3.187 \times 10^{-5}$ percent of $Gs$.\par

Seeking to implement our proposed strategy, we developed our own sparse solver to compute semidefinite programs. Although it can only calculate results for small cases, we believe it to be a stepping stone in the right direction.\par
In order to create our own solver, we had to utilize Python’s pre-existing linear algebra tools as the tools that [6, 9] employs within its solvers,BLAS or Lapack, use dense matrices. Additionally, our solving method switched from the more specific semidefinite programming technique to the broader cone-solving one.
Implementing our sparse solver, we began by:
\begin{lstlisting}
Gsp = sp.coo_matrix((list(G.V), (list(G.I), list(G.J))),
	(Nsd, c.size[0]))
Asp = sp.coo_matrix((list(A.V), (list(A.I), list(A.J))),
	((kQ-1)+kQ*nrindep,num_variables()))
\end{lstlisting}

Here, we take the matrices $G$ and $A$, and transform them into the sparse matrices: Gsp and Asp. \par
We then created a set of functions which implement a series of linear algebraic functions on Gsp and Asp as outlined by the quantum algorithm:
\begin{lstlisting}
WWT = matrix(0.0, (cdim, cdim))
for i in xrange(cdim):
     WWT[i,i] = -1.0
     misc.scale(WWT, W, trans = 'N', inverse = 'N')
     misc.scale(WWT, W, trans = 'T', inverse = 'N')
     wwt = sp.coo_matrix(WWT)
     lin = sp.bmat([[None, Asp.transpose(), Gsp.transpose()], [Asp, None, None],
     [Gsp, None, wwt]]) 

\end{lstlisting}

After coding the constraints of $A*x=b$ and $Gs[k]*x + ss[k] = hs[k]$ to finish off our solver, we were able to compute a result using:

\begin{lstlisting}
solution = solvers.conelp(c, G, h, dims, A = A, b = b, kktsolver=kkt_sp
	(G, dims, A))
\end{lstlisting}

We then computed the case $(S(2,3))$ for both our sparse and initials solvers. Our results show that the sparse solver took .00907 seconds on average to solve the problem, while the dense one took .05936 seconds. These results show that the sparse solver took 84\% less time than the initial one to solve the SDP problem; although we understand that this is a small case, it highlights the future time improvements possible for larger cases which our solver can not currently solve.

\section{Conclusion and Future Work}
We formulated a technique that will further the solutions of the SDP using the notion of sparse matrices. As our sparse program is not currently optimized, this is an area of future work as one can seek to create a fully sparse solver which can compute the SDP for even larger cases. We believe this will allow us to get closer to the lower bound of $(\ln N - 1)/\pi$ and thus be a great improvement to the field of Quantum Complexity Theory.


\begin{thebibliography}{9}
\bibitem{1} E. Farhi, J. Goldstone, S. Gutmann, and M. Sipser, MIT CTP \#2815, (1999), arXiv:quant-ph/9901059.

\bibitem{2} A. Childs, A. Landahl, and P. Parrilo, Phys. Rev. A \textbf{75}, 032335 (2007), arXiv:quant-ph/0608161v1

\bibitem{3} P. H{\o}yer, J. Neerbek, and Y. Shi, Algorithmica \textbf{34}, 429 (2002), arXiv:quant-ph/0009032.

\bibitem{4} L. Fej´er, J. Reine Angew. Math. \textbf{146}, 52 (1915).

\bibitem{5} F. Riesz, J. Reine Angew. Math. \textbf{146}, 83 (1915).

\bibitem{6} M. Andersen, L. Vandenberghe, \emph{CVXOPT},
available from http://cvxopt.org/copyright.html

\bibitem{7} B.Jacokes,\emph{An Improved Quantum Algorithm for Searching an Ordered List}
http://web.mit.edu/rsi/www/pdfs/papers/2003/2003-brianj.pdf

\bibitem{8} M. Nielsen and I.L. Chuang, \emph{Quantum Computation and Quantum Information}, Cambridge University Press (2000)

\bibitem{9} NumPy, http://www.numpy.org/
		 	 	 		
\bibitem{10} L. K. Grover, in /emph{Proceedings of the Twenty-Eighth Annual ACM Symposium on Theory of Computing}, edited by Gary L.Miller ACM, New York, 1996 , pp. 212–219. 
					 				
			
\end{thebibliography}
\end{document}